\documentclass{ws-procs11x85}

\def\Journal#1#2#3#4{{#1} {\bf #2}, #3 (#4)}

\makeindex
\begin{document}

\title{WIMP/Neutralino Direct Detection}

\author{M. DE JESUS}

\address{IPN Lyon-UCBL, IN2P3-CNRS, 4 rue Enrico Fermi, 69622 Villeurbanne
Cedex, France \\E-mail: dejesus@in2p3.fr}



\twocolumn[\maketitle\abstract{   The most popular
candidate for non baryonic dark matter is the neutralino. More than twenty experiments are
dedicated to its direct detection. This review describes the most competitive and promising   experiments with different detection
techniques. The most recent results are presented with some
prospects for the near future. }]

\baselineskip=13.07pt
\section{Introduction}

The existence of dark matter in the Universe is now  well established  in the
astro-particle community reenforced by the recent astrophysical observations of the satellite experiment
WMAP\cite{wmap} : about 27\% of the mass-energy of the Universe is composed of matter.  
Ordinary matter (baryons) contributes to about 4\% of this total mass density  of the Universe and
only $\simeq$ 1 \% is  visible according to the most recent measurements of
the amount of deuterium in high red-shift clouds of gas and of the CMB\cite{gas}.  
Hence about  90\% of this dark matter is not baryonic.  We have to distinguish two categories, hot and cold dark matter
particles refering to their velocity at the matter-radiation decoupling time in the early
Universe.  Hot dark matter implies moving relativistically and cold  moving non-relativistically. Neutrinos
with  non-zero masses are hot dark matter candidates, however WMAP\cite{wmap} results  combined with other
experiments and observations lead to a contribution $<1.5 \%$ for light neutrino species. 

So the bulk of the non-baryonic dark matter is cold dark matter (CDM). Among the numerous solutions proposed by theorists 
axions and neutralinos are favorites. Neutralinos are candidates of the generic class of Weakly Interacting Massive 
Particles (WIMP).
Axion  are particles  proposed to solve the strong CP violation problem in the Peccei-Quinn theory
\cite{peccei}. Astrophysical considerations combined with experimental
constraints require an axion mass in the range $10^{-3}$ to $10^{-6} eV/c^2$ . For a more detailed
discussion about axions see reference\cite{axion1,axion2,axion3}, since this paper will be dedicated to the
WIMP/neutralino detection.
The neutralino is the lightest supersymmetric particle, a linear
combination of the supersymmetric partners of the photon, Z and Higgs bosons, in
the minimal~supersymmetric~extension~of~the~standard~model :
\begin{equation}
\chi^0 = a \tilde \gamma + b \tilde Z + c \tilde{H_1^0} + d \tilde{H_2^0}
\label{eq:neutralino}
\end{equation}

 Its mass is constrained to lie in the range 45~GeV~$< m_{\chi} <$~3~TeV, where the lower bound comes from 
accelerator results from LEP and the upper bound is given by  astrophysical constraints such as the age
of the Universe or unitarity. 
Locally our galaxy is supposed to be imbedded in a WIMP halo. 
 
Many experiments are  dedicated to direct and indirect detection of WIMPs, two complementary
techniques. Direct detection experiments measure the energy deposited by elastic scattering of 
a neutralino of our own galaxy off a target nucleus.  For  masses larger than $\simeq$ 200 GeV,
indirect detection of dark matter 
particles through their annihilation products may be more suitable. In this paper we will concentrate on  
the case of direct detection techniques, and for a complete description of  indirect approaches we refer to the
papers\cite{indirect1}$^-$\cite{indirect6}.

In the direct detection approach the expected event rate depends on various parameters coming from
astrophysics,
particle physics and nuclear physics; it can range from 1 to $10^{-5}$ events/kg/day. The measured signal is very 
low (few keV) depending  on the masses of the
incident particle and of the scattered nucleus, but also on the nuclear recoil relative efficiency
(quenching factor) in producing charges, light or heat. Hence  WIMP direct searches 
put strong constraints on  experimental background environments, and require detectors with very low energy 
thresholds.
In this review we present the different possible signatures for disentangling  a WIMP signal from the
background. Different experimental approaches are described and illustrated  by a few
experiments. The current limits in the exclusion plot and near future prospects will be also 
presented.

\section {WIMP/neutralino direct detection physics principles}

\begin{figure}
\center
\psfig{figure=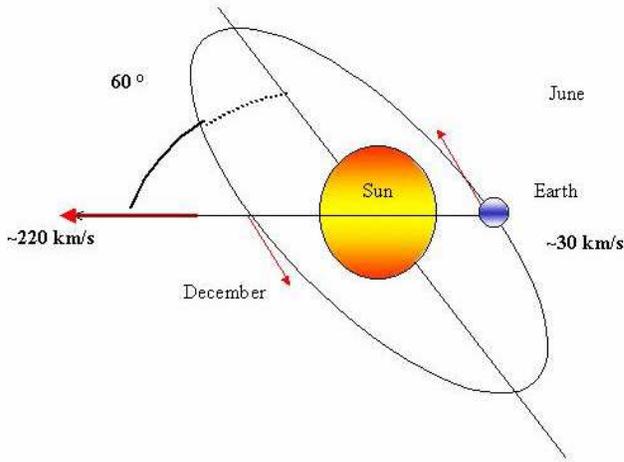,width=10.0truecm}
\caption{Annual modulation.}
\label{fig:modann}
\end{figure}

As mentionned previously, WIMP-nucleus interaction rate depends on various parameters. First we have to
define a WIMP halo model. For simplicity the approximation of a maxwellian velocity 
distribution in the galactic frame is made (see \cite{kamionkowski} for a review on alternative halo models).
Next, a supersymmetric model is chosen for predicting the WIMP interaction with quarks of nucleons inside the 
target nucleus. Depending on the chosen model  the WIMP-nucleus cross-section has two components\cite{smith,jungman} :  spin-dependent and  spin-independent. 
The spin-independent term couples to the mass of the nucleus and the spin-dependent couples to
its spin. The nuclear form factor depends on  the nature of the interaction. The spin-dependent case is
the most complicated one, requiring detailed nuclear models (for more details see
dedicated papers\cite{jungman,ressel}). In the following we will restrict this review to the simplest 
spin-independent case which is supposed to dominate in most models for massive target nuclei.
Taking into account these previous considerations the interaction rate can be expressed as follows :

\begin{equation}
{{dR}\over {dQ}} = {{\sigma_0 \rho_h}\over {2 m^2_r m_\chi}} F^2(Q) {\int_{v_{min}}^\infty} {f(v)\over v} dv
\label{eq:rate}
\end{equation}

where $m_r$ is the WIMP-nucleus reduced mass $m_\chi m_N/(m_\chi+m_N)$, $m_\chi$ is the WIMP mass, $m_N$ is the
nucleus mass.

$\rho_h = 0.3 GeV/c^2/cm^3$ is the assumed halo WIMP density at the position of the solar system, 
$f(v)$ is the dark matter velocity distribution,  with an average rms velocity 
$v_0 = 220 km/s$,  truncated above the escape velocity of the galaxie $v_{esc}\simeq 575 km/s$,
$\sigma_0$ is the total nucleus-WIMP interaction cross section and $F(Q)$ is the nuclear form factor. 

\subsection{Exclusion plot $\sigma (m_{\chi})$}

In order to reliably compare supersymmetric models with results obtained by different experiments 
using different techniques a $\sigma_0 (m_{\chi})$ plot is built in the following way. The cross-sections $\sigma_0 (m_{\chi})$ are normalized to a single 
nucleon $\sigma (m_{\chi})$ to allow comparisons between  different target nuclei.
The measured nuclear recoil event rate is compared to a
theoretical spectrum calculated for a given  WIMP mass and cross-section.
If an experiment observes a signal then we build a $\sigma (m_{\chi})$ contour plot. 
If the observed events cannot be unambigously associated with a WIMP signal an exclusion limit is calculated.
WIMP signals have distinctive signatures that backgrounds are not supposed  able to mimic. Three different 
signatures are proposed.

\begin{figure}
\center
\psfig{figure=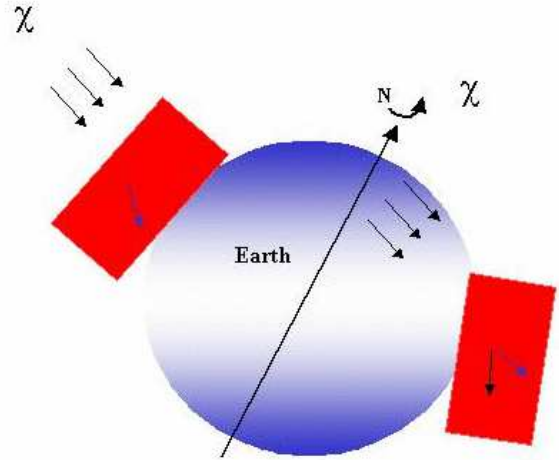,width=8.0truecm}
\caption{Diurnal modulation.}
\label{fig:moddiu}
\end{figure}

\subsection{Annual modulation }\label{subsec:modann}

As a result of the   Earth motion around the Sun the count rate in detectors should show an annual
modulation (fig.~\ref {fig:modann}). Along the year the Earth's velocity relative to the galaxy varies, as in June the Earth and the Sun
velocities add up  whereas in December they subtract\cite{hasen98}. The maximum amplitude of this effect in 
the signal is about 7\%.  We will
report  later that the DAMA collaboration using NaI scintillating crystals is the first experiment  and at the moment 
the only one, claiming for an evidence of a WIMP annual modulation signal.

\subsection{Diurnal modulation and directionnality}\label{subsec:moddiu}

Another possible modulation in the WIMP signal is the night and day variation, this effect is due to the shielding of the
detector by the 
Earth of the incident flux. For masses close to 50 GeV and under certain
assumptions the diurnal modulation can be larger than the annual one\cite{spergel88}. However the most interesting daily signature
coupled with the annual one  is the directionnality of the WIMP wind as illustrated in
figure~\ref{fig:moddiu}. This  effect is also larger than the annual one. The validation of the principle has been performed by the DRIFT-I experiment with a 1 m$^3$ low 
pressure TPC\cite{martoff00} prototype.

\subsection{Target atomic mass effect}

Observed together annual and diurnal modulations are unambigous methods to distinguish WIMP and background signals, but they are
very difficult to operate. In the spin-independent case, an easiest method is to use different target materials as the event rate depends on
the target atomic mass. To give an estimate of this effect we can use the Smith and Lewin\cite{smith} calculated 
integral rate $R_0$ with no form-factor correction
and an average recoil energy $E_{R}$

\begin{equation}
R_0 \simeq 5.87\times  v_0 \sigma_{m_\chi } \rho_h A^3 {{m_\chi }\over{(m_A+m_\chi)^2}} /kg/d
\label{eq:rzero}
\end{equation}

\begin{equation}
E_R \simeq 2\times 10^{-6} m_\chi^2 ({v_0 \over c})^2  {{m_A }\over{(m_A+m_\chi)^2}} keV
\label{eq:er}
\end{equation}

Table ~\ref{tab:targets} reports on  $R_0$ and $E_{R}$ values for different targets, and for
a given WIMP mass of $m_\chi \simeq 50$ GeV, $\sigma_{m_\chi } \simeq 7.10^{-6}pb$. Naively if we consider the event rate it
seems to be more advantageous to use high mass nuclei, but if we look to the recoil energy
as the target atomic number A increases, the average deposited energy tends to decrease. So the choice of a target is a
compromise between these two quantities. Moreover we can see for example that germanium is more efficient than silicon for WIMPs
detection while they have similar cross-sections for neutrons.\par
 Another important point is the possible neutron multiple
scattering in the detector, which is  impossible for a WIMP. We will see hereafter this  method is used by
 the CDMS collaboration\cite{akerib1}, with germanium and silicon targets as illustrated in 
 figure~\ref{fig:cdms}.

\begin{table}
\begin{center}
\caption{Integrated event rate $R_0$ and average energy deposition for different
target atomic masses, no form-factor correction and $m_\chi
= 50$ GeV, $v_0 = 220 km/s$ and $\sigma _{n\chi} = 7\times 10^{-6}$ pb.\label{tab:targets}} \vspace{0.2cm}
\begin{tabular}{|c|c|c|c|}
\hline
\raisebox{0pt}[12pt][6pt]{} &
\raisebox{0pt}[12pt][6pt]{$A$} &
\raisebox{0pt}[12pt][6pt]{$R_0$} &
\raisebox{0pt}[12pt][6pt]{$<E_R>$} \\
\hline
\raisebox{0pt}[12pt][6pt]{H} &
\raisebox{0pt}[12pt][6pt]{1} &
\raisebox{0pt}[12pt][6pt]{5.10$^{-5}$} &
\raisebox{0pt}[12pt][6pt]{1} \\
\hline
\raisebox{0pt}[12pt][6pt]{Na} &
\raisebox{0pt}[12pt][6pt]{23} &
\raisebox{0pt}[12pt][6pt]{0.3} &
\raisebox{0pt}[12pt][6pt]{11} \\
\hline
\raisebox{0pt}[12pt][6pt]{Si} &
\raisebox{0pt}[12pt][6pt]{28} &
\raisebox{0pt}[12pt][6pt]{0.5} &
\raisebox{0pt}[12pt][6pt]{12} \\
\hline
\raisebox{0pt}[12pt][6pt]{Ge} &
\raisebox{0pt}[12pt][6pt]{73} &
\raisebox{0pt}[12pt][6pt]{3} &
\raisebox{0pt}[12pt][6pt]{13} \\
\hline
\raisebox{0pt}[12pt][6pt]{I} &
\raisebox{0pt}[12pt][6pt]{127} &
\raisebox{0pt}[12pt][6pt]{8} &
\raisebox{0pt}[12pt][6pt]{11} \\
\hline
\raisebox{0pt}[12pt][6pt]{Xe} &
\raisebox{0pt}[12pt][6pt]{131} &
\raisebox{0pt}[12pt][6pt]{9} &
\raisebox{0pt}[12pt][6pt]{11} \\
\hline
\raisebox{0pt}[12pt][6pt]{Pb} &
\raisebox{0pt}[12pt][6pt]{210} &
\raisebox{0pt}[12pt][6pt]{18} &
\raisebox{0pt}[12pt][6pt]{8} \\
\hline
\end{tabular}
\end{center}
\end{table}

\section {WIMP/neutralino direct detection techniques}

\begin{figure*}
\psfig{figure=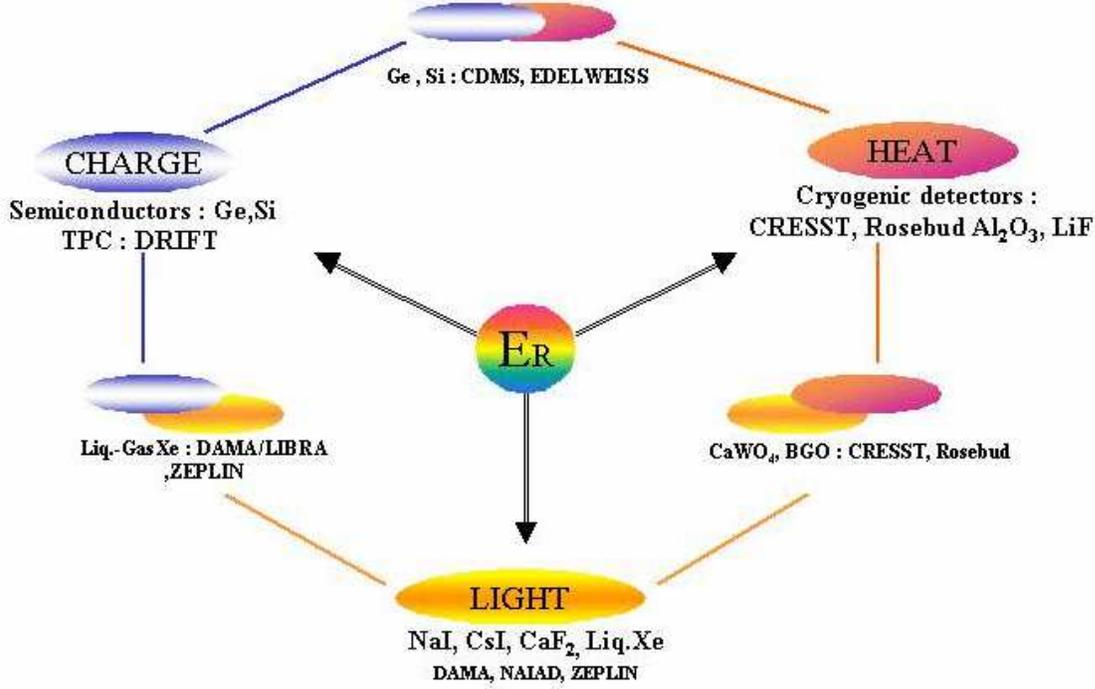,width=6truein}
\caption{Illustration of the different techniques developped for the WIMP direct detection.}
\label{fig:dephys} 
\end{figure*}

WIMP  detectors  are constrained by  three important requirements : low threshold, ultra low 
background and high mass detector.
When a WIMP interacts with a nucleus, the nuclear recoil can induce different signals (fig.
~\ref{fig:dephys}) : heat, ionization and scintillation. During the last decade important technical 
developments were based on one or two of  these different physics processes.

\subsection{Quenching factor}

A relevant parameter in WIMP direct detection is the relative efficiency of nuclear recoil   called 
quenching factor. It is the ratio of the number of charge carriers produced by a nuclear 
recoil due to the WIMP interaction over an electron recoil of the same kinetic
energy (electron equivalent energy or "eee"). For scintillating materials  the quenching factor is defined as the ratio between the light 
produced by a nuclear  recoil and by an electron recoil. \par
\noindent While in conventional detectors this factor is usually below 30\% (measured, e.g. to be
$\simeq$ 0.3 for germanium\cite{messous}, $\simeq$ 0.25 for sodium and $\simeq$0.08 for 
iodine\cite{gerbier} ), for cryogenic detectors described hereafter it has been measured to be around 
one for  recoiling nuclei independently on energy\cite{zhou,alessandrello,sicane}.

\subsection{Classical detectors : semiconductors and scintillators}

Germanium diodes initially used in double beta decay experiments were the first detectors  used to search  for WIMPs, 
since they have very low thresholds and very good resolutions. 
Experiments like IGEX\cite{igex1,igex2} and HDMS\cite{hdms}, with about 2 kg 
of enriched $^{76}Ge$,  achieved very low background count
rates ($<$0.2 evt/kg/day in the interval 10-40 keV) and $E_{thr} \simeq 4-10$ keV-ee 
(equivalent to $\simeq 15-30$ keV recoil).

\begin{figure}
\center
\psfig{figure=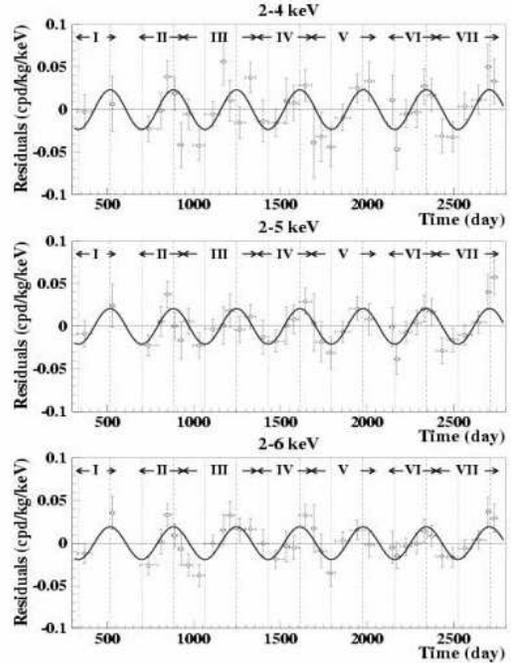,width=7.0truecm}
\caption{DAMA model independent residual count rates as a function of time for 7 years and three energy intervals (2-4), (2-5)
and (2-6) keV-ee.
\label{fig:damamod}}
\end{figure}

\begin{figure}
\center
\psfig{figure=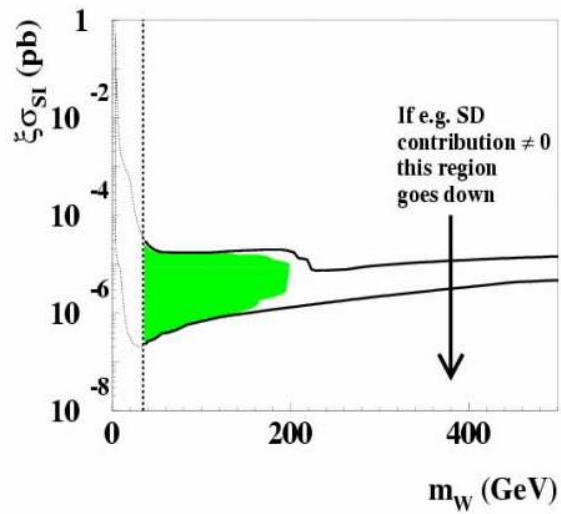,width=8.0truecm}
\caption{Dama limit for the dominant spin-independent case obtain with 7 years of data 
taking. This contour plot is obtain with different WIMP-halo models, see
ref$^{29}$ for a detailed discussion.}
\label{fig:damalimit}
\end{figure}

\begin{figure}
\center
\psfig{figure=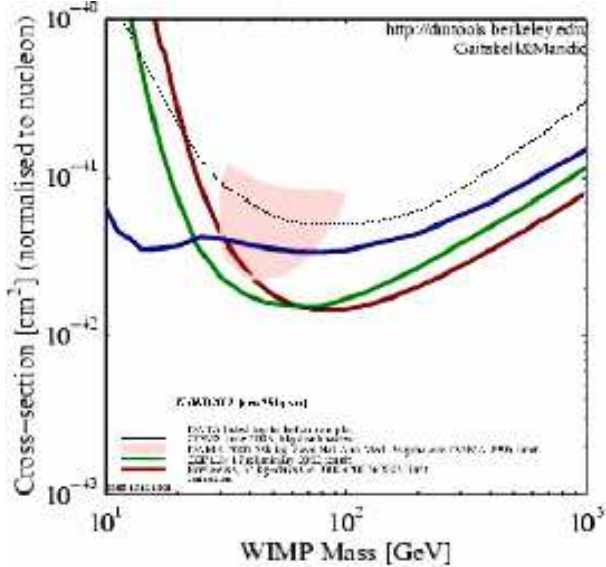,width=8.0truecm}
\caption{Current spin-independent limits for the most competitive experiments. The WIMP halo parameters used are $\rho_h=0.3
GeV/c^{-2}cm^{-3}$, $v_0=220 km/s$. The closed contour corresponds to the 3$\sigma$ allowed region of the DAMA first four
years obtain with the same WIMP halo parameters.}
\label{fig:limit03}
\end{figure}

Large masses were easily achievable with scintillators like NaI or liquid xenon in a very pure environment. The DAMA
experiment has operated more than 100 kg of NaI (each crystal weighting about 9.7 kg with energy threshold of 
$\simeq 2 keV-ee$ ie 22 keV recoil) for several years in the Gran Sasso underground laboratory. They accumulated data
during 7 years and since 1997 they announce evidence for an annual modulated WIMP signal. The DAMA group claim 
their observation  is compatible with a signal induced by a WIMP of $\simeq 52$ GeV mass
ad a WIMP-nucleon cross section of $\simeq 7.2 $ pb.  
The DAMA collaboration has published\cite{dama03} this last summer the last 3 years campaign totalizing 7
years and confirms their observation of an annual modulation signal as illustrated in 
figs.~\ref{fig:damamod},\ref{fig:damalimit}. Right now none of the currently running dark matter  experiments confirms this
signal as we can see in the current exclusion plot in fig.~\ref{fig:limit03}. Independent experiments with NaI detectors 
(NAIAD\cite{naiad}
in the Boulby mine, ANAIS\cite{anais} in Canfranc, ELEGANT\cite{elegant} in Oto Cosmo Observatory) are currently running.  
The NAIAD\cite{naiad1} experiment most recent results begin to exclude   the DAMA $\sigma(m_\chi)$ region in the
spin-independent exclusion plot.

As we have seen previously despite the very high purity level of classical detectors,  they suffer ultimately from a lack of 
power discrimination between electron and nuclear recoils.

The first discrimination method used   is based on a pulse shape analysis.
It is a statistical method where the measured quantity is the rise-time of the light signal which depends on the
nature of the recoiling particle. This discrimination method is used with sodium iodide crystals 
(DAMA, NAIAD) but is also successfully used with liquid scintillators like liquid xenon. \par
With a 3.1 kg liquid-Xenon detector the ZEPLIN-I\cite{sumner}
collaboration has reached preliminary sensitivities which could exclude the DAMA zone. However some problems remain : a
relativily high electronic background rate has to be understood, there's no nuclear recoil calibration for the low 
energy part of the spectrum ($<$50 keV-ee), a poor energy resolution compared to bolometers. Some of these points should
be answered in the next few months as the experiment in now  currently running deep underground in the BOULBY mine\cite{smith}. \par
The DAMA/LIBRA collaboration is currently running a new NaI detector mith a larger mass ($\approx$
250~kg) as well as  a liquid-Xenon detector.

The future projects ZEPLIN-II  and -III aimed to be able to develop a discrimination technique with
a two phase  liquid-gas Xenon detector with charge and light signals.  

\begin{figure}
\center
\psfig{figure=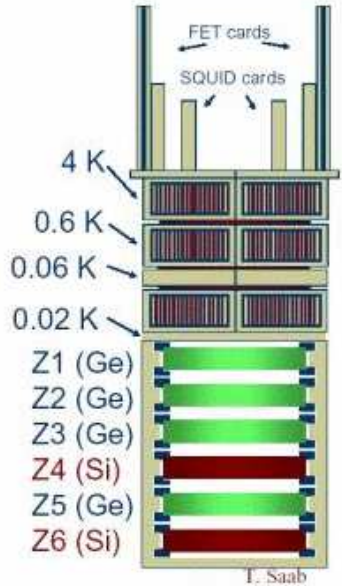,width=5.0truecm}
\caption{CDMS detector tower.}
\label{fig:cdms}
\end{figure}

\subsection{Cryogenic detectors}

Since the beginning of the 90's important developments were also made in new directions like cryogenic 
detectors. They are made of a crystal with a thermometer glued on it, operating  at very low temperature 
(few tens of millikelvin). Very low thresholds were reached  by the CRESST-I experiment\cite{cresst1} with a 262 g sapphire 
calorimeter (resolutions of $\simeq 133$ eV at 1.5 keV and thresholds $\simeq 500$ eV).   \par
 But most impressive results were obtained with mixed techniques allowing the simultaneous measurement of two components
heat-light or heat-charge.  The two combined informations are a powerful tool to distinguish
 a nuclear recoil induced by a WIMP or a neutron interaction from electron recoils induced by a gamma
 or an electron interaction (quenching factor described previously). 
It is an event by event discrimination method. Again different approaches were explored by different worldwide collaborations.
 For cryogenic detectors the CDMS and EDELWEISS
collaborations investigate the heat-ionization way, and  the CRESST and ROSEBUD collaborations explore the
heat-light channels.

The CDMS collaboration was the first\cite{cdms0,cdms01} to operate a detector giving simultaneously ionization and heat
 signals with a germanium crystal. 
Until 2002 the experiment was running in the shallow  site in Stanford with a poor muon shielding  inducing an important 
 neutron background. Despite this limitation they 
derive competitive dark matter limits and were leaders for several years. They could  subtract the neutron 
backgroung using a monte carlo simulation but also taking advantage of the fact that they run simultaneously two 
different targets :
germanium and silicon\cite{akerib1,akerib2}. 
During the year 2003 the CDMS-II experiment is being installed in  the deep underground Soundan mine  where the muon
flux is reduced by 5 orders of magnitude reducing the neutron background by a factor 400. 
They are currently operating 2 towers (fig.~\ref{fig:cdms}) of 3x165 g Ge and 3x100 g Si detectors  and 18 more detectors are under
fabrication totalizing 4 kg of germanium. The CDMS collaboration expects to improve its current sensitivity ($\simeq
1 evt/kg/day$) by two orders of magnitude.

\begin{figure}
\center
\psfig{figure=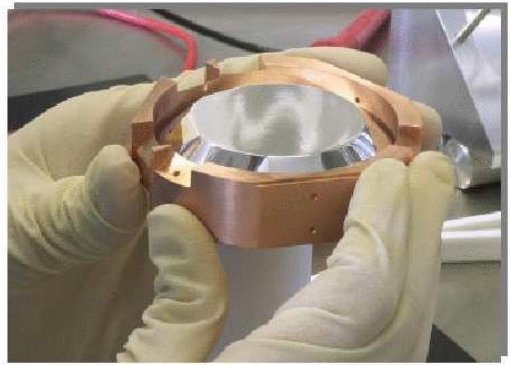,width=7.0truecm}
\caption{EDELWEISS 320 g Ge detector.}
\label{fig:edw}
\end{figure}

\begin{figure}
\center
\psfig{figure=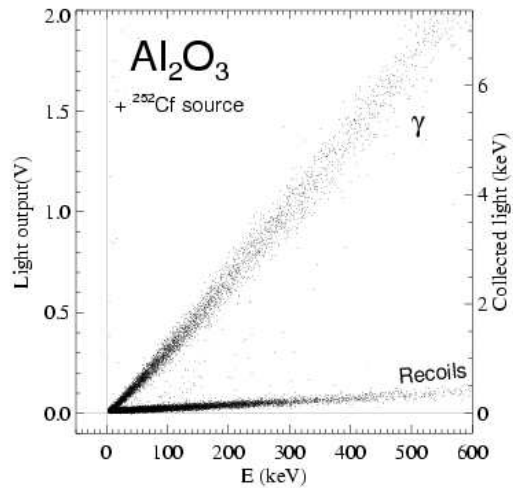,width=7.0truecm}
\caption{Discrimination between gammas and nuclear recoils in a 50 g sapphire bolometer at 20 mK by the  ROSEBUD
collaboration .}
\label{fig:rosebud}
\end{figure}

The currently best spin-independent
published limit was obtained by the EDELWEISS collaboration cumulating 32 kg.d. The EDELWEISS experiment is
 installed in the underground laboratory of Modane in the French-Italian Alps. They operate similar detectors to those of CDMS 
germanium crystals (fig.~\ref{fig:edw})  with different technologies for the electrodes\cite{edw1} runing at 
$\simeq 18 mK$. Three 320 g detectors are running simultaneously.
During the last campaign in june 2003, 2 events were observed in the nuclear recoil zone which origin is
 under investigation.  
More data is being analysed, but the EDELWEISS-I  stage data taking will be soon  finished. For the next stage a larger cryostat with a detection volume of 100 litres is built  and is
currently beeing tested. 
This cryostat benefits from an original technology developped at the CRTBT-Grenoble 
laboratory. 
 The EDELWEISS-II installation will take a year from now.
 The first step will operate 21x320 g germanium detectors with NTD thermometers and 7x200 g NbSi thin
film germanium detectors developped by the group of the CSNSM laboratory\cite{csnsm}. 
A muon veto made of 140 $m^2$ plastic scintillator will be  added. It should reject the  neutron background
induced by cosmic muons in the inner lead shielding, which has been  evaluated  two orders of magnitude below the present EDELWEISS-I 
sensitivity $\simeq 0.2$ evt/kg/day.  Such  background has to be clearly identified and rejected since the expected  event rate for the 
EDELWEISS-II stage is about $10^{-2}$ evt/kg/day. 
In a second step up to 120 detectors will operate simultaneously

The CRESST-II\cite{cresst2} and ROSEBUD\cite{demarcillac} experiments involve scintillating crystals as cryogenic detectors. They operate in the same way ; 
the heat is measured with thermometer glued on the scintillator and the light is collected with a second thin but large
 surface crystal. The main advantage of such method is the large possibility for scintillating target materials : CaWO$_4$,
 PbWO$_4$, Al$_2$O$_3$, BaF, BGO, ... and for important volumes.
 A few years ago S.P\'ecourt et al.\cite{sophie} characterized the phonon channel of a 1 kg Al$_2$O$_3$ bolometer and
 recently the same team\cite{demarcillac} has succeeded in measuring the light output of a 50 g Al$_2$O$_3$
 bolometer(fig.~\ref{fig:rosebud}). \par
 The CRESST-II\cite{cresst2} experiment should operate 33x300 g modules of CaWO$_4$ totalizing about 10 kg.

\subsection{New promising techniques}

\begin{figure}
\center
\psfig{figure=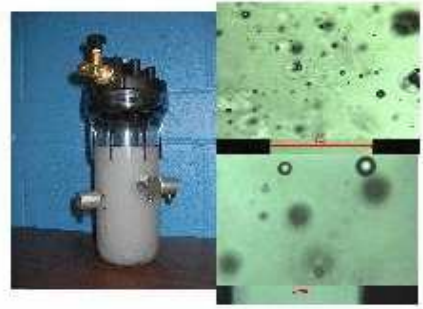,width=6.0truecm}
\caption{PICASSO new 1liter module }
\label{fig:picasso}
\end{figure}

\begin{figure}
\center
\psfig{figure=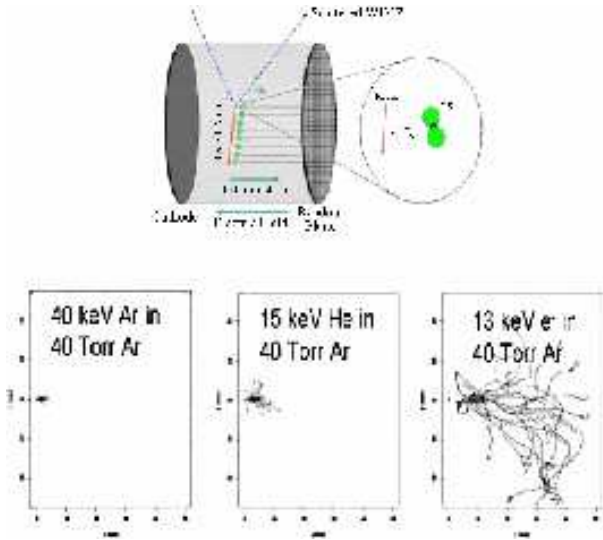,width=8.0truecm}
\caption{DRIFT-1 ionization tracks for three different types of recoiling particles : argon, helium and electrons}
\label{fig:drift}
\end{figure}

In addition to the techniques described above, illustrated by currently running experiments and their
near future, other promising techniques are under investigation. 

The PICASSO\cite{picasso1,picasso2} and SIMPLE\cite{simple} experiments have choosen to adapt a well known technology used in
neutron dosimetry, to develop a counter for WIMP induced nuclear recoils. The method is based on small superheated Freon
droplets imbedded in a gel matrix at room temperature. The nuclear recoil of $^{19}F$ induces the explosion of a droplet , creating an acoustic shock
wave measured with  piezoelectric transducers.   By varying the temperature of the gel the energy threshold can be triggered in such a way that the
electron recoil induced by gamma background can be supressed. 
Calibration is made at different pressures and temperatures with
monoenergetic neutrons produced by a Van de Graff Tandem . The use of $^{19}F$ (spin-$1/2$ isotope) is particularly interesting to search for
spin-dependent neutralinos. A first generation of detectors, 16 modules of  8 ml, lead to the
published limit of the PICASSO collaboration\cite{picasso1,picasso2}. They are currently
running the second generation of modules with a larger volume (fig.~\ref{fig:picasso})) in an improved low background
environnement in the SNO underground laboratory : PIC@SNO. New purification techniques were
developped especially for the PICASSO experiment \cite{picasso3}. Despite a very good backgroung discrimination the main disadvantage of 
such an integrating  detector is the necessity to run the experiment at different threshold energies in order
 to measure the deposited energy spectrum. \par

To take advantage of the directionnality which appears as the clearest signature of WIMPs, the UKDMC collaboraton has developped 
and is currently running successfully, the DRIFT-I detector. 
It consists in  a 1 m$^3$ low pressure TPC filled with a $Xe-CS_2$ gas mixture. The principle of the TPC is well known, 
the innovation is the use of $CS_2$ negative ions instead of 
$e^-$ as charge carriers reducing the diffusion in order to achieve millimetric 
track resolution (fig.~\ref{fig:drift}). Important  improvements on the read-out techniques such as 
MICROMEGAS\cite{luscher}, in
order to increase the pressure hence the target mass, are underway. Other possible target gases are also studied to
prepare the next generations DRIFT-II and -III with a larger gas mass  for the TPC of the order of 100 kg. 

\begin{figure}
\psfig{figure=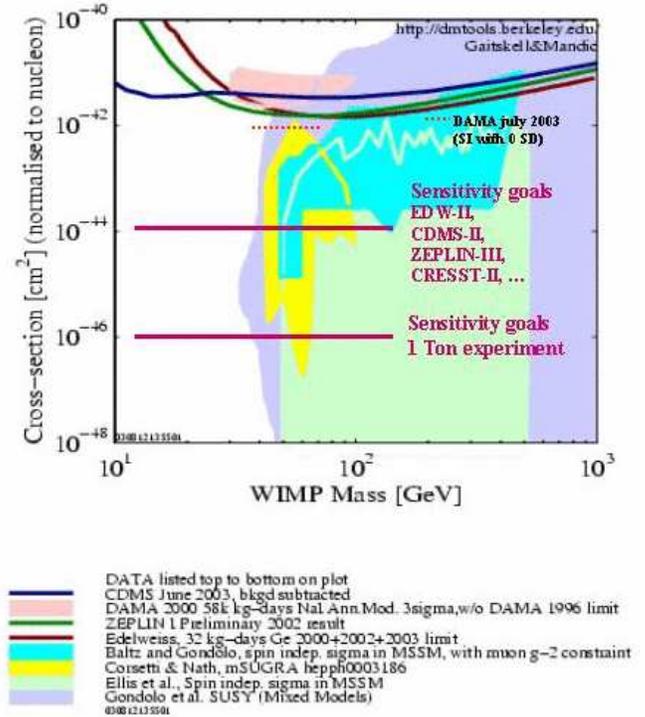,width=3.5truein}
\caption{Projected limits for some of the next generation experiments. The colored regions represent different SuSy model 
calculations 
\label{fig:prospect} }
\end{figure}

\section{Conclusions}
The current experimental spin-independent limit turns  around 10$^{-6}$~pb which corresponds to 
a count rate of about 0.2 to 
1~evt/kg/day. To achieve this limit it took about 10 years for most of the
currently running first generation experiments to develop these detectors. 
The next generation under construction and for most of them on the final stage, aim to improve this limit by two orders of magnitude, that means a count 
rate around 10$^{-2}$~evt/kg/day. This has a price : lowering the sensitivity by about two orders of magnitude  implies increasing 
the
target mass by about the same factor (for example EDELWEISS-I worked with 3x320 g Ge and EDELWEISS-II should run at
the end 120x320 g Ge
detectors).\par
With this scaling the ultimate  neutron background induced by muons can no longer be neglected. It is the reason why 
experiments like EDELWEISS-II,CDMS-II and CRESST-II will use a muon veto. \par
The next five years are very promising : a clarification of the DAMA annual modulation signal is essential. Indirect
Earth-based and Space experiments like Antares, HESS, AMS and GLAST should give  independent cross checks.
Meanwhile accelerator physics will explore an important part of SuSy space parameters on the exclusion plot (fig.~\ref{fig:prospect}).
\par 
Nevertheless the one-tonne scale experiment will probably involve larger  international collaborations. The technical
challenge will be to build an experiment able to achieve the extremely low background necessary to cover most of the
prediction mSUGRA models.

\section*{Acknowledgments}
I would like to thank the in2p3 for its financial support.

\end{document}